# Demonstrating sub-3 ps temporal resolution in a superconducting nanowire single-photon detector

B. A. Korzh[1,a),b)], Q-Y. Zhao[2,b)], S. Frasca[1], J. P. Allmaras[1,3], T. M. Autry[4], E. A. Bersin[1,2], M. Colangelo[2], G. M. Crouch[1], A. E. Dane[2], T. Gerrits[4], F. Marsili[1], G. Moody[4], E. Ramirez[1], J. D. Rezac[4], M. J. Stevens[4], E. E. Wollman[1], D. Zhu[2], P. D. Hale[4], K. L. Silverman[4], R. P. Mirin[4], S. W. Nam[4], M. D. Shaw[1] and K. K. Berggren[2]

[1]*Jet Propulsion Laboratory, California Institute of Technology, Pasadena, California 91109, USA*

[2]*Department of Electrical Engineering and Computer Science, Massachusetts Institute of Technology, Cambridge, Massachusetts 02139, USA*

[3]*Applied Physics, California Institute of Technology, Pasadena, California 91125, USA*

[4]*National Institute of Standards and Technology, Boulder, Colorado 80305, USA*

**Improving the temporal resolution of single photon detectors has an impact on many applications[1], such as increased data rates and transmission distances for both classical[2] and quantum[3–5] optical communication systems, higher spatial resolution in laser ranging and observation of shorter-lived fluorophores in biomedical imaging[6]. In recent years, superconducting nanowire single-photon detectors[7,8] (SNSPDs) have emerged as the highest efficiency time-resolving single-photon counting detectors available in the near infrared[9]. As the detection mechanism in SNSPDs occurs on picosecond time scales[10], SNSPDs have been demonstrated with exquisite temporal resolution below 15 ps[11–15]. We reduce this value to 2.7±0.2 ps at 400 nm and 4.6±0.2 ps at 1550 nm, using a specialized niobium nitride (NbN) SNSPD. The observed photon-energy dependence of the temporal resolution and detection latency suggests that intrinsic effects make a significant contribution.**

Temporal resolution in SNSPDs, commonly referred to as jitter, is characterized by the width of the temporal distribution of signal outputs with respect to the photon arrival times. This statistical distribution is known as the instrument response function (IRF), and its width is commonly evaluated as

---

a) Electronic mail: boris.a.korzh@jpl.nasa.gov

b) These authors contributed equally to this work.

the full-width at half-maximum (FWHM). Significant effort has been made in the community to reduce the jitter and to ultimately understand the fundamental limitations. Though timing fluctuations may arise from the microscopic physics of the detection mechanism, only limited indications of this have been observed to date, such as asymmetry in the IRF[16–18]. This is primarily because jitter in SNSPDs has been limited by instrumental mechanisms, which has prevented a systematic experimental study of the intrinsic timing characteristics on different device design parameters. Resolving the intrinsic jitter of SNSPDs is essential for the validation of a microscopic model of the detection mechanism, an active topic of research in the community[10].

The main effect that masks the intrinsic jitter in SNSPDs is noise jitter, which arises due to amplifier noise-induced shifts of the readout signal that lead to fluctuations in the threshold crossing time. Noise jitter can be reduced by engineering devices with higher operating currents and by using lower-noise radio-frequency (RF) cryogenic amplifiers, which has led to the current state-of-the-art jitter of 14.2 ps (see Ref.[14] for a review of recent efforts). Despite this progress, a conclusive demonstration of intrinsic effects is still missing, typically evidenced by a lack of systematic temperature or nanowire-width dependence of the measured jitter[13]. In addition, it was recently realized that fluctuations in the longitudinal position of the photon absorption lead to significant changes in the detection latency (defined as the time between the absorption of a photon and the registration of an electrical output pulse) since the high kinetic inductance of typical nanowire structures renders the speed of RF pulse propagation along the wire to be a small fraction of the speed of light in vacuum[19]. Fortunately, this "geometric" jitter can be compensated, at least in part, by using a differential readout scheme with a low-noise amplifier at each end of the nanowire[20].

In this Letter, we describe specialized NbN SNSPD devices (illustrated in Figure 1) designed to simultaneously minimize the noise jitter and the geometric jitter at the expense of system detection efficiency. This allows us to carry out a comprehensive study, demonstrating significant photon-energy, nanowire-width and temperature dependence of the jitter, as well intrinsic variations in the detection latency. These results make an important step towards probing the fundamental physics of the detection process and can be used to validate theoretical models of the detection mechanism and predictions of jitter in SNSPDs. In addition, they show that SNSPDs can outperform any other free-running single-photon detection technologies in terms of jitter[1,3]: we observed a jitter of 2.7 ps for visible wavelengths, whereas the lowest jitter achieved by micro-channel plate photomultiplier tubes[21] and Si avalanche diodes[22] is 20 ps. In the near-infrared, our results show 4.6 ps jitter and the best alternative detectors are InGaAs avalanche diodes, that are capable of 50 ps jitter[23]. We note that although sub-picosecond jitter and single-photon sensitivity are achievable with commercial streak cameras, they are incompatible with applications requiring free-running operation or high count rates, which SNSPDs can provide.

To put these results into context, one can consider the potential impact on a handful of applications. The combination of an SNSPD with a time-correlated single-photon counting (TCSPC) module results in an optical sampling oscilloscope[6], where an IRF width of 2.7 ps is equivalent to a signal bandwidth of about 130 GHz, comparable to the fastest available sampling oscilloscopes. Crucially, the sensitivity and dynamic range of such an SNSPD-based instrument could be many orders of magnitude greater than an oscilloscope based on a fast photodiode. Laser ranging would also benefit from our fast SNSPD: detection of a single visible photon reflected from an object, could yield a spatial precision of 1 mm with a confidence of approximately 90%. Averaging of multiple detections adds further advantage, with a resolution of 10 μm possible after just 10,000 events[6] - a task easily completed on a millisecond time scale if a TCSPC module is used[12]. In the context of quantum key distribution (QKD)[4], jitter of single-photon detectors sets the maximum system clock rate, influencing the effective signal-to-noise ratio and in turn setting the maximum allowable channel loss[23]. Separating the optical pulses in a QKD system by the width of the detector IRF at 1/100 of the maximum gives a base error rate of 1%; with our detector this would allow a clock rate of up to 80 GHz, more than 1.5 orders of magnitude higher than that used in the record distance QKD demonstration[5].

The IRF for one of the devices tested is shown in Figure 2(a), demonstrating a clear dependence on the detected wavelength. With 1550 nm light, we observed 4.6±0.2 ps jitter, which is reduced to 3.8±0.3 ps for 775 nm light, at the same bias current. Since the 775 nm and 1550 nm pulses were synchronized in time (within a small offset, see Methods for details), we were also able to observe a measurable difference in the relative detection latency between the two wavelengths. This has previously not been observable, since SNSPD jitter values were too large to distinguish small latency changes. We observe that detection signals resulting from 1550 nm photons are delayed by 4.2±0.4 ps relative to 775 nm photons at a bias current of 21.5 µA. Reducing the bias current increases both the jitter and the relative detection latency, as illustrated in Figure 2(b).

To illustrate the overall effect of varying the nanowire width and photon energy, we plot the jitter as a function of bias current for these variables in Figure 3(a). This plot shows that for a fixed bias current, the jitter decreases for narrower nanowires and for higher energy photons. The same trend is evident in the normalized photon count rates (PCR) shown in Figure 3(b): for a given bias current, increasing the photon energy or reducing the nanowire width leads to higher internal detection efficiency. The existence of saturated PCR plateaus suggests that the devices are operating with near-unity internal detection efficiency at high bias currents[9]. Increasing the photon energy increases the length of the plateau, and our results suggest that using a detector with a long plateau is the key to achieving low intrinsic jitter, which has not been understood previously.

Due to the indication that increasing the photon energy reduces the jitter, we measured the widest nanowire (120 nm), using shorter wavelengths, down to 273 nm, as illustrated in Figure 4(a). The widest nanowire was chosen because it showed the largest jitter for a given bias current, suggesting the largest contribution of intrinsic effects and the best opportunity to observe a significant photon-energy dependence. As before, the jitter curves are correlated with the PCR curves as illustrated in Figure 4(b). At the highest bias current, the jitter decreased from 5.9±0.3 ps for 1550 nm to 3.2±0.3 ps for visible and ultra-violet wavelengths. In this measurement, we noticed a saturation of the jitter for high photon energies and bias current. (see caption of Figure 4 for discussion). with Figure 4(c) showing the IRF for illumination with 400 nm. The most significant effect of photon energy is observed at low bias currents, with the jitter changing from 24.8±1.8 ps for 1550 nm to 6.6±0.6 ps for blue and ultra-violet light at around 18.3 μA.

In addition to the photon-energy dependence presented here, we also observe a weak temperature dependence as discussed in Supplementary Note 4. This, along with the nanowire-width dependence, suggests that intrinsic mechanisms due to device physics dominate the jitter, indicating an avenue for future investigations of the fundamental physics of SNSPDs[10]. We note that recently published work indicated a photon-energy dependence of the jitter[17]. However, because the experiment was carried out on flood-illuminated meandered structures, it is difficult to make comparisons with theoretical predictions due to additional contributions from geometric effects as well as detections in the turns (Note that high efficiency meandered SNSPDs are typically designed to avoid detections in the turns[9]). Our work also reveals that the detection latency depends on the photon energy, with Figure 3(c) showing the detection latency difference between 775 nm and 1550 nm photons, for all devices presented. This data can be used as an additional tool in validating microscopic theories of the photon detection process. Reproducing the energy-dependent detection latency differences could prove more straightforward than comparing the jitter to microscopic models, since many sources of instrumental fluctuations and timing offsets can be eliminated. Furthermore, a model could average over transverse coordinate dependence[24], Fano fluctuations[25], and thermal fluctuations, reducing the complexity and computational demands of generating a full IRF for model validation. We note that recent theoretical studies have outlined a method for simulating the detection latency in SNSPDs[26], however, comparison with our experimental results would require dedicated numerical simulations that are outside the scope of this Letter. An energy-dependent detection latency could also have practical implications, since it could open avenues to engineering devices with coarse energy or photon-number resolution in clocked systems, which has previously not been possible with single-pixel SNSPDs.

We have demonstrated jitter as low as 2.7±0.2 ps for 400 nm and 4.6±0.2 ps at 1550 nm, using a specialized NbN SNSPD, which is considerably lower than competing technologies. This was achieved by eliminating the effect of geometric jitter by keeping the devices short, reducing the noise jitter through the use of a low-noise cryogenic amplifier, as well as achieving devices with high internal-detection efficiency. While the small size of the devices results in low absorption probability, the techniques shown here could be applied toward practical low-jitter devices with high detection efficiency, either by using a differential readout to remove the geometric jitter[20] or by integrating the detector with a photonic waveguide[11] or field-enhancement cavity[27]. The observed saturation of jitter at short wavelengths (see Figure 4(a)), which is likely due to limits of the characterization setup as well as the noise floor of our readout scheme, suggests that there is still room for further optimization of the noise jitter and that the intrinsic jitter could be even lower than 2 ps. This optimization could be achieved using a near quantum-limited amplifier[28], a superconducting digital readout element such as a nanocryotron[29], by using an adiabatic taper to impedance match the nanowire to 50 $\Omega$[19], or by engineering SNSPDs with faster rise times[30]. Our investigation has taken an important step towards better understanding the fundamental origin of timing jitter in SNSPDs and calls for further theoretical investigations, which could lead to improved performance in the future.

## METHODS

**Nanofabrication and screening.** The devices were patterned from a NbN film with a nominal thickness of 7 nm. We deposited the film in a reactive sputtering system, nominally at room temperature on a 4 inch silicon wafer with a 300 nm-thick thermal oxide layer[31]. As deposited, the film sheet resistance was $R_S = 340\ \Omega/\square$ at room temperature and the residual resistance ratio was $\rho_{RRR} = 0.8$. The critical temperature of the film was measured to be $T_c = 8.65$ K. The entire device was patterned with one step of electron-beam lithography. We spun an 80 nm-thick positive-tone e-beam resist (GL2000[32], diluted to 5%) baked at 180°C for 2 minutes. The e-beam lithography was performed with a 125 kV system. We separated the writing into two steps. The first step exposed the nanowire, which was designed as a coplanar waveguide structure. To expose the gap area of the nanowire with fine edge roughness, the exposure beam current was 500 pA with a step size of 1 nm. The second step was to expose the outline of the electrical contacts. Because the size of the contacts exceeded the size of the maximum writing field, we applied a multi-pass method with a 10 nA beam current and a step size of 4 nm. We developed the chip in O-xylene at 5°C for 30 seconds and a subsequent rinse in isopropanol for another 30 seconds. To transfer the pattern to the NbN film, we performed a $CF_4$ reactive ion etch at 50 W for 3 minutes. After the etch, resist was applied to protect the device from oxidization and dicing. We first removed the residual GL-2000 layer in 60°C N-methyl pyrodone for 10 minutes, and then spun a new layer of GL-2000 and a 1.5 µm-thick photoresist (Microposit S1813[32]) for protection. After dicing, the S1813 was stripped with acetone while the GL-2000 layer was left as a protection layer. It was possible to wire bond though the GL-2000 layer directly to the NbN contact pad.

The total length of the inductor was 1.5 mm, corresponding to an inductance of 96 nH assuming a sheet inductance of 64 pH/$\square$, extracted from pulse shape measurements in SNSPD structures fabricated from similar films. We applied a hyperbolic curve to the two ends of the short nanowire to minimize current crowding[33] and reduce the length of the taper area.

Overall, 160 devices were fabricated on 20 dies in a single fabrication run. Approximately a third of the devices were screened to measure the switching currents at 0.9 K, and the PCR curves at 1550 nm were measured for about 20 of those screened. This allowed us to select a representative device for each of the four nanowire widths for tests with the low-noise cryogenic amplifier.

**Cryogenic setup.** The experiment was carried out by using a pulse-tube cryocooler with a $^4$He sorption refrigerator at a base temperature of 0.9 K. The signal from the SNSPD was amplified with a SiGe cryogenic amplifier (Cosmic Microwave, CITLF1[32]) mounted on the 4 K stage of the cryocooler. The amplifier had a nominal gain of 50 dB, a nominal bandwidth of 1.5 GHz with a low frequency cut-off at 1 MHz and a nominal noise temperature of 7 K. The SNSPD was biased with a low-noise current source through a 5 kΩ resistive bias-T at the input of the amplifier. The amplifier was biased with a supply voltage of 3 V and dissipated approximately 30 mW of power.

**Optical setup.** The initial characterization of the jitter was carried out using a mode-locked fiber laser with a center wavelength of 1550 nm and a nominal pulse width of 0.5 ps. A synchronization signal was generated by splitting a fraction of the 1550 nm light with a fiber coupler and sending it onto a fast photodetector. We verified that the SNSPDs were operating in the single-photon detection regime by characterizing the detection rate as a function of incident optical power at different bias currents and operating in the range yielding linear dependence.

To investigate the energy dependence of the jitter and detection latency, we used a 0.5 mm-long, periodically poled lithium niobate (PPLN) second harmonic generation (SHG) crystal to frequency double the mode-locked laser to 775 nm. After the crystal, the light was collimated and free-space coupled into the cryostat through a series of glass windows in the vacuum chamber and the heat shields at 40 K and 4 K, flood illuminating the device under test. The polarization of the light was orientated parallel to the nanowire (TE polarization). The optical intensity was controlled with a circular metallic variable neutral-density filter. This configuration ensured that the converted 775 nm light and the unconverted 1550 nm light co-propagated via the same path through the optical setup. It was calculated that the 775 nm pulse was delayed by 1.1±0.1 ps, relative to the 1550 nm light, due to chromatic dispersion in the optical elements after the SHG crystal. All detection latency results presented in this Letter have been corrected for this offset. After generation, filters were used to select the 1550 nm or 775 nm wavelength illumination.

The measurements with a broader range of wavelengths were carried out with an optical setup based on a tunable titanium-sapphire mode-locked laser capable of pumping an optical parametric oscillator (OPO), configurable for visible and near-infrared operation. The center wavelength of the laser was set to 800 nm, which resulted in a pulse width of 130 fs at a repetition rate of 76 MHz. The OPO was used to generate 600 nm and 1200 nm light with a pulse duration of 200 fs. Pulses at 400 nm were achieved by frequency doubling the pump laser, with a type-I β-Barium borate (BBO) crystal. 273 nm light was generated with a BBO-based tripler with the pump laser tuned to 820 nm.

**Jitter and latency measurement.** The SNSPD and laser synchronization signals were acquired simultaneously on a digital real-time oscilloscope with a sampling rate of 40 Giga samples per second and the time delay between the two pulses was recorded for each acquisition. Histograms of 20,000 detection delays were collected for each jitter measurement, yielding the IRF, which typically required a collection time of approximately 5 minutes. The optimum analog bandwidth setting of the oscilloscope was found to be 6 GHz, which is sufficiently high for the SNSPD signal which had a rise time (20% to 80%) of approximately 80 ps. Increasing the bandwidth further generates additional noise, which increases the noise jitter. The vertical scale setting of the oscilloscope was also optimized in order to minimize the noise floor of the input whilst limiting saturation effects. To investigate the effect of the oscilloscope sample rate, we carried out one measurement with an oscilloscope capable of 80 Giga samples per second (see Figure 4(c)) which reduced the lowest measurable jitter from 3.2 ps to 2.7 ps. The reason for this reduction is due to an improved reproduction of both the SNSPD and synchronizations signals. Since the laser synchronization signal had a rise time of approximately 50 ps, it is actually under-sampled with a rate of 40 Giga samples per second, resulting in amplitude and timing fluctuations of the acquired waveform.

It is difficult to obtain the absolute detection latency due to the complexity of calibrating the exact moment of photon absorption in the nanowire. Instead, we investigated the detection latency difference between two photon energies. Since the 775 nm and 1550 nm pulses are synchronized in our optical setup (see above) an IRF was collected for each bias current with both wavelengths without realigning the beam, simply by replacing the free-space filter. The trigger level was kept the same for the two wavelengths at each bias current, and it was verified that the SNSPD pulse shape was the same for both wavelengths (see Supplementary Note 2). The latency difference was calculated by measuring the shift of the 1550 nm data relative to 775 nm, taking the peak of the fitted IRF as the reference.

**Instrument response function (IRF) fitting.** It has been recognized by several groups that the IRF of SNSPDs, in certain regimes, is non-Gaussian, exhibiting a significant tail at long delays[16–18]. This was clearly observed in our data. We found that an exponentially modified Gaussian (EMG)[34] distribution fit the experimental data much more closely than a normal distribution, a method that has also been adopted elsewhere[16,17]. An EMG function[34] is defined as

$$y = \lambda e^{\{\sigma^2 \lambda^2 /2 - \lambda(t-\mu)\}} \Phi\left\{\frac{t-\mu}{\sigma} - \sigma\lambda\right\},$$

where $\sigma$ is the standard deviation of the normal distribution, $\lambda$ is the exponential decay rate, $\mu$ is the average of the distribution and $\Phi$ is the normal cumulative distribution function. This function considers the sum of independent normal and exponential random variables. The nonlinear least-squares fitting routine in MATLAB[32] was used to fit the EMG to the data. The EMG fit parameters for all devices presented in this Letter are outlined in Supplementary Note 1. Error values presented throughout this Letter correspond to 95% confidence bounds of the resulting IRF fit.

# REFERENCES

1. Eisaman, M. D., Fan, J., Migdall, a & Polyakov, S. V. Invited review article: Single-photon sources and detectors. *Rev. Sci. Instrum.* **82,** 71101 (2011).

2. Hemmati, H. *Deep Space Optical Communications*. (2006).

3. Hadfield, R. H. Single-photon detectors for optical quantum information applications. *Nat. Photonics* **3,** 696–705 (2009).

4. Lo, H.-K., Curty, M. & Tamaki, K. Secure Quantum Key Distribution. *Nat. Photonics* **8,** 595–604 (2015).

5. Korzh, B. *et al.* Provably secure and practical quantum key distribution over 307 km of optical fibre. *Nat. Photonics* **9,** 163–168 (2015).

6. Becker, W. *Advanced Time-Correlated Single Photon Counting Techniques*. (2005). doi:10.1007/3-540-28882-1

7. Gol'tsman, G. N. *et al.* Picosecond superconducting single-photon optical detector. *Appl. Phys. Lett.* **79,** 705–707 (2001).

8. Natarajan, C. M., Tanner, M. G. & Hadfield, R. H. Superconducting nanowire single-photon detectors: physics and applications. *Supercond. Sci. Technol.* **25,** 63001 (2012).

9. Marsili, F. *et al.* Detecting single infrared photons with 93 % system efficiency. *Nat. Photonics* **7,** 210–214 (2013).

10. Engel, A., Renema, J. J., Il'in, K. & Semenov, A. Detection mechanism of superconducting nanowire single-photon detectors. *Supercond. Sci. Technol.* **28,** 114003 (2015).

11. Pernice, W. H. P. *et al.* High-speed and high-efficiency travelling wave single-photon detectors embedded in nanophotonic circuits. *Nat. Commun.* **3,** 1325 (2012).

12. Shcheslavskiy, V. *et al.* Ultrafast time measurements by time-correlated single photon counting coupled with superconducting single photon detector. *Rev. Sci. Instrum.* **87,** 53117 (2016).

13. You, L. *et al.* Jitter analysis of a superconducting nanowire single photon detector. *AIP Adv.* **3,** 72135 (2013).

14. Wu, J. *et al.* Improving the timing jitter of a superconducting nanowire single-photon detection system. *Appl. Opt.* **56,** 2195–2200 (2017).

15. Esmaeil Zadeh, I. *et al.* Single-photon detectors combining high efficiency, high detection rates, and ultra-high timing resolution. *APL Photonics* **2,** 111301 (2017).

16. Najafi, F. Timing performance of Superconducing Nanowire Single-Photon Detectors. (2015).

17. Sidorova, M. *et al.* Physical mechanisms of timing jitter in photon detection by current-carrying superconducting nanowires. *Phys. Rev. B* **96,** 184504 (2017).

18. Caloz, M. *et al.* High-detection efficiency and low-timing jitter with amorphous superconducting nanowire single-photon detectors. *Appl. Phys. Lett.* **112,** 61103 (2018).

19. Zhao, Q.-Y. *et al.* Single-photon imager based on a superconducting nanowire delay line. *Nat. Photonics* **11,** 247–251 (2017).

20. Calandri, N., Zhao, Q. Y., Zhu, D., Dane, A. & Berggren, K. K. Superconducting nanowire detector jitter limited by detector geometry. *Appl. Phys. Lett.* **109,** 152601 (2016).

21. Kume, H., Koyama, K., Nakatsugawa, K., Suzuki, S. & Fatlowitz, D. Ultrafast microchannel plate photomultipliers. *Appl. Opt.* **27,** 1170–8 (1988).

22. Cova, S., Lacaita, A., Ghioni, M., Ripamonti, G. & Louis, T. A. 20-Ps Timing Resolution With Single-Photon Avalanche Diodes. *Rev. Sci. Instrum.* **60,** 1104–1110 (1989).

23. Amri, E., Boso, G., Korzh, B. & Zbinden, H. Temporal jitter in free-running InGaAs/InP single-photon avalanche detectors. *Opt. Lett.* **41,** 5728 (2016).

24. Wu, H., Gu, C., Cheng, Y. & Hu, X. Vortex-crossing-induced timing jitter of superconducting nanowire single-photon detectors. *Appl. Phys. Lett.* **111,** 62603 (2017).

25. Kozorezov, A. G. *et al.* Fano fluctuations in superconducting-nanowire single-photon detectors. *Phys. Rev. B* **96,** 54507 (2017).

26. Vodolazov, D. Y. Single-Photon Detection by a Dirty Current-Carrying Superconducting Strip Based on the Kinetic-Equation Approach. *Phys. Rev. Appl.* **7,** 34014 (2017).

27. Akhlaghi, M. K., Schelew, E. & Young, J. F. Waveguide integrated superconducting single-photon detectors implemented as near-perfect absorbers of coherent radiation. *Nat. Commun.* **6,** 8233 (2015).

28. Ho Eom, B., Day, P. K., LeDuc, H. G. & Zmuidzinas, J. A wideband, low-noise superconducting amplifier with high dynamic range. *Nat. Phys.* **8,** 623–627 (2012).

29. McCaughan, A. N. & Berggren, K. K. A superconducting-nanowire three-terminal electrothermal device. *Nano Lett.* **14,** 5748–53 (2014).

30. Smirnov, K. V. *et al.* Rise time of voltage pulses in NbN superconducting single photon detectors. *Appl. Phys. Lett.* **109,** 52601 (2016).

31. Dane, A. E. *et al.* Bias sputtered NbN and superconducting nanowire devices. *Appl. Phys. Lett.* **111,** 122601 (2017).

32. The use of trade names is intended to allow the measurements to be appropriately interpreted and does not imply endorsement by the US government, nor does it imply these are necessarily the best available for the purpose used here.

33. Clem, J. R. & Berggren, K. K. Geometry-dependent critical currents in superconducting nanocircuits. *Phys. Rev. B* **84,** 174510 (2011).

34. Grushka, E. Characterization of exponentially modified Gaussian peaks in chromatography. *Anal. Chem.* **44,** 1733–1738 (1972).

35. Kerman, A., Yang, J., Molnar, R., Dauler, E. & Berggren, K. Electrothermal feedback in superconducting nanowire single-photon detectors. *Phys. Rev. B* **79,** 100509 (2009).

## ACKNOWLEDGMENTS

Part of the research was performed at the Jet Propulsion Laboratory, California Institute of Technology, under contract with the National Aeronautics and Space Administration. Support for this work was provided in part by the DARPA Defense Sciences Office, through the DETECT program and the National Science Foundation under grant number ECCS-1509486. E.A.B., A.E.D., G.M.C. and J.P.A. acknowledge partial support from the NASA Space Technology Research Fellowship program. E.R. acknowledges support from the MARC-U*STAR program. D.Z. acknowledges support from the A*STAR National Science Scholarship. We thank Alexander Kozorezov, Andrew Beyer, Peter Day, Varun Verma, Billy Putnam, Daniel Santavicca, and Bill Rippard for valuable discussions and loan of measurement equipment.
## AUTHOR CONTRIBUTIONS

B.A.K., Q.-Y.Z. conceived and designed the experiments. B.A.K., Q.-Y.Z., S.F., E.R., E.A.B., M.J.S., T.M.A., G.M., M.C., D.Z. and A.E.D. performed the experiments. B.A.K., S.F. and J.P.A. analyzed the data. B.A.K., Q.-Y.Z., S.F., E.A.B., T.G., M.J.S., T.M.A., G.M., M.C., D.Z., A.E.D., E.E.W., G.M.C., J.P.A., J.R., P.D.H, K.L.S., R.P.M., S.W.N., F.M., M.D.S and K.K.B contributed materials/analysis tools. B.A.K., Q.-Y.Z. and M.D.S wrote the paper with input from all authors.

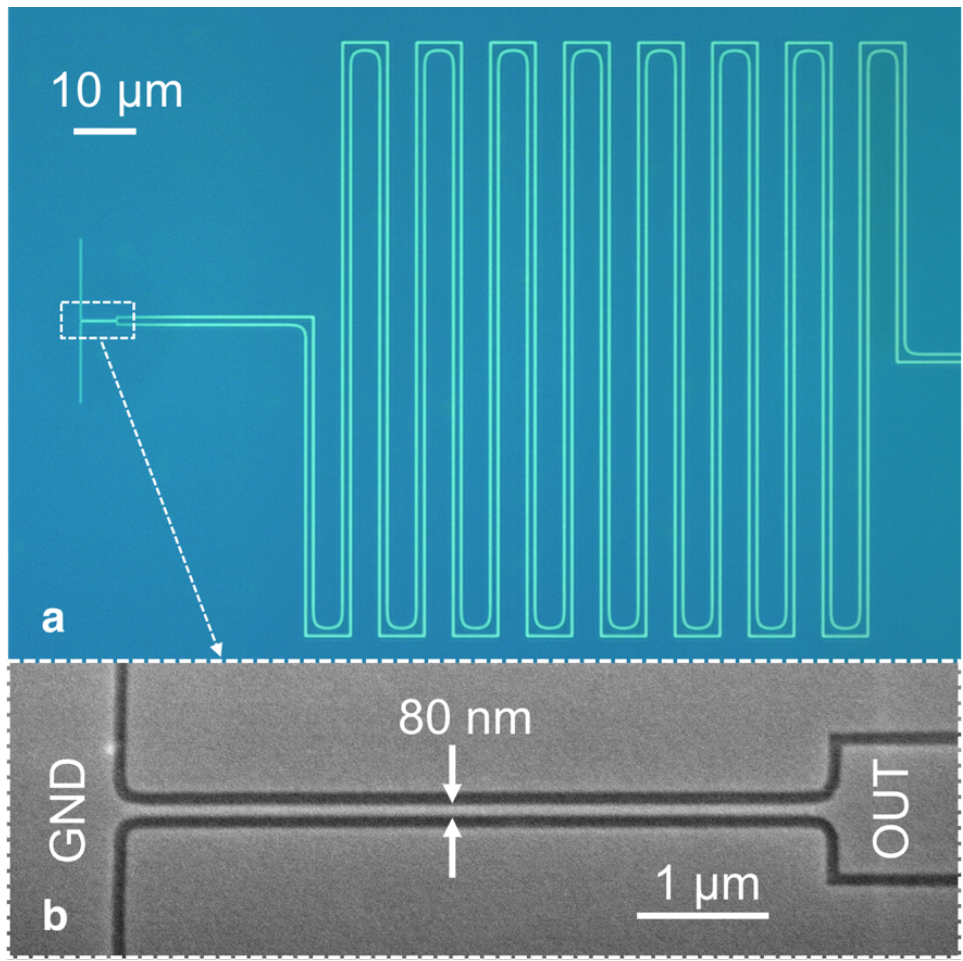

**Figure 1 | Low-jitter superconducting nanowire single-photon detector.** (a) Optical micrograph of a representative device, with a small dashed rectangle on the left showing the nanowire region. Darker blue represents the NbN, while the lighter green/blue color represents the regions removed in the $CF_4$ etch. The large meandering structure on the right is an inductor designed to prevent the short device from latching[35] and is wide enough (1 µm) to prevent it from being single photon detector itself. (b) Scanning electron micrograph of the active area. Only the 80 nm-wide nanowire is displayed here, however, four different widths were studied in this work: 60 nm, 80 nm, 100 nm, and 120 nm. The nanowire length of 5 µm length, which is short relative to standard devices, was selected to keep the predicted geometric contribution to the jitter below 1 ps, assuming that the speed of signal propagation in the nanowire is approximately 2% of the speed of light in vacuum[19].

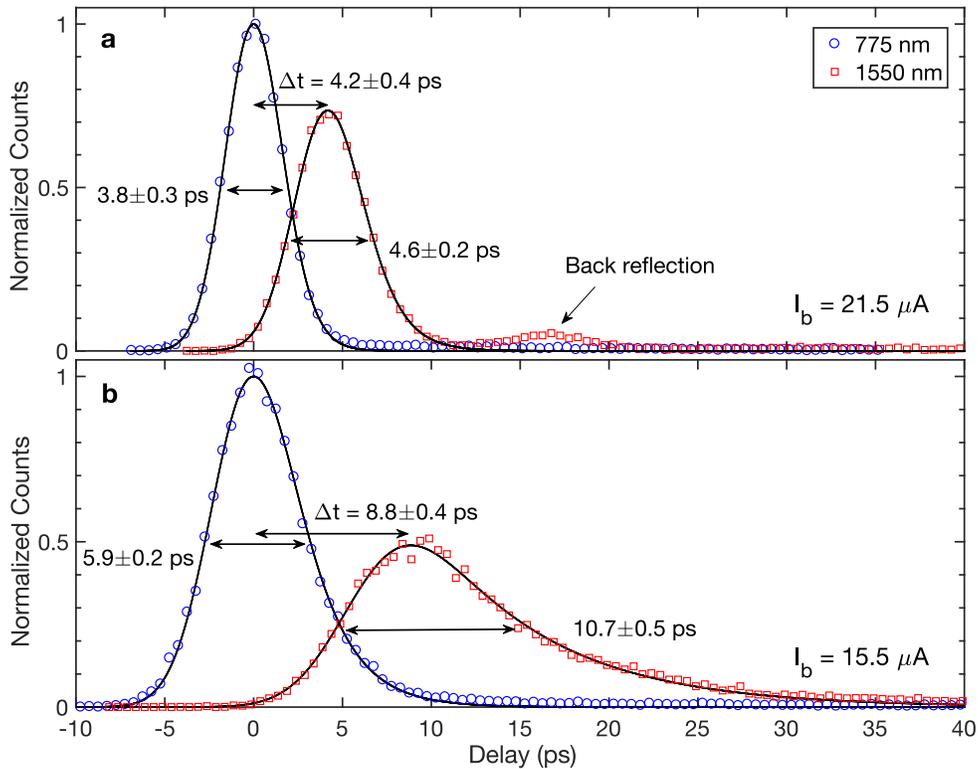

**Figure 2 | Instrument response function showing wavelength dependence of detection latency and jitter.** IRF with illumination at 775 nm (blue circles) and 1550 nm (red squares) wavelengths at bias currents of (a) 21.5 µA and (b) 15.5 µA for the 80 nm-wide nanowire. Note that the histograms show a significant exponential component at lower bias, and are dominated by the Gaussian component at high bias (see Methods for description of fitting function and Supplementary Note 1 for the fitting parameters). The relative latency between the two wavelengths increases at lower bias currents. The fact that both the jitter and latency depend on the wavelength suggests that we are probing the intrinsic detection mechanism. (a) The 1550 nm data shows a second peak at a delay of 12.4±0.4 ps, with an amplitude of 6% of the main peak. This can be explained by a reflection from the back of the Si wafer, which was 500 µm thick. This peak was not observed in the 775 nm data, which is consistent with this explanation since silicon is not transparent at this wavelength. It will be possible to reduce or eliminate this peak in the future by embedding the 1550 nm devices in a broadband optical stack, or by anti-reflection coating

the back side of the wafer. The FWHM values are indicated, as well as the relative latency between the 775 nm and 1550 nm detections, calculated from the peaks of the IRFs.

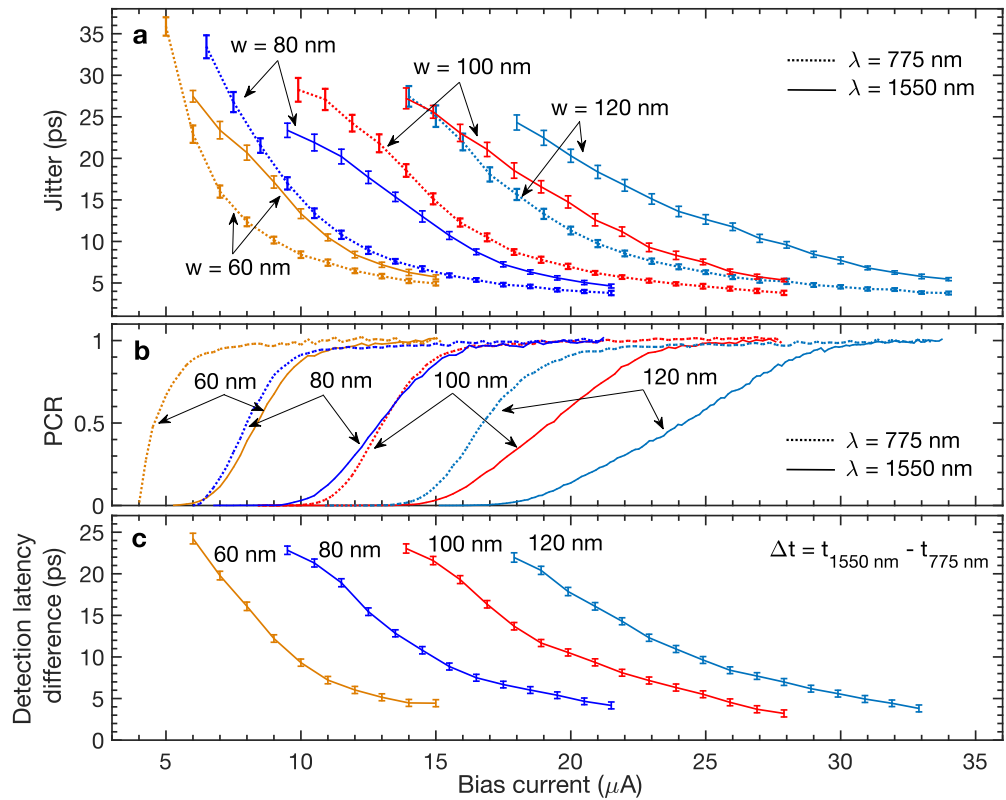

**Figure 3 | Bias current dependence of the jitter, normalized photon count rate (PCR) and detection latency difference.** Jitter (a) and normalized PCR (b) as a function of bias current, for illumination at 775 nm (dashed lines) and 1550 nm (solid lines). Nanowire widths are represented with different colors: 60 nm (orange), 80 nm (blue), 100 nm (red), and 120 nm (pink). The switching currents of the devices at 0.9 K ranged between 16 µA for the 60 nm-wide nanowire and 35 µA for the widest nanowire. We found that for 1550 nm light, the 80 nm device, which had a switching current of around 22 µA, exhibited the lowest jitter of 4.6±0.2 ps. This data suggest a correlation between the length of the saturation plateau and the minimum achievable jitter. (c) Bias-current dependence of the mean detection latency difference between the detection of a 775 nm photon and a 1550 nm photon. Positive values indicate that detections due to 1550 nm photons are delayed relative to 775 nm detections. The error bars in (a) represent the 95% confidence bounds of the IRF fit and in (b) they represent the uncertainty in the relative arrival times of the 1550 nm and 775 nm pulses at the detector (see Methods).

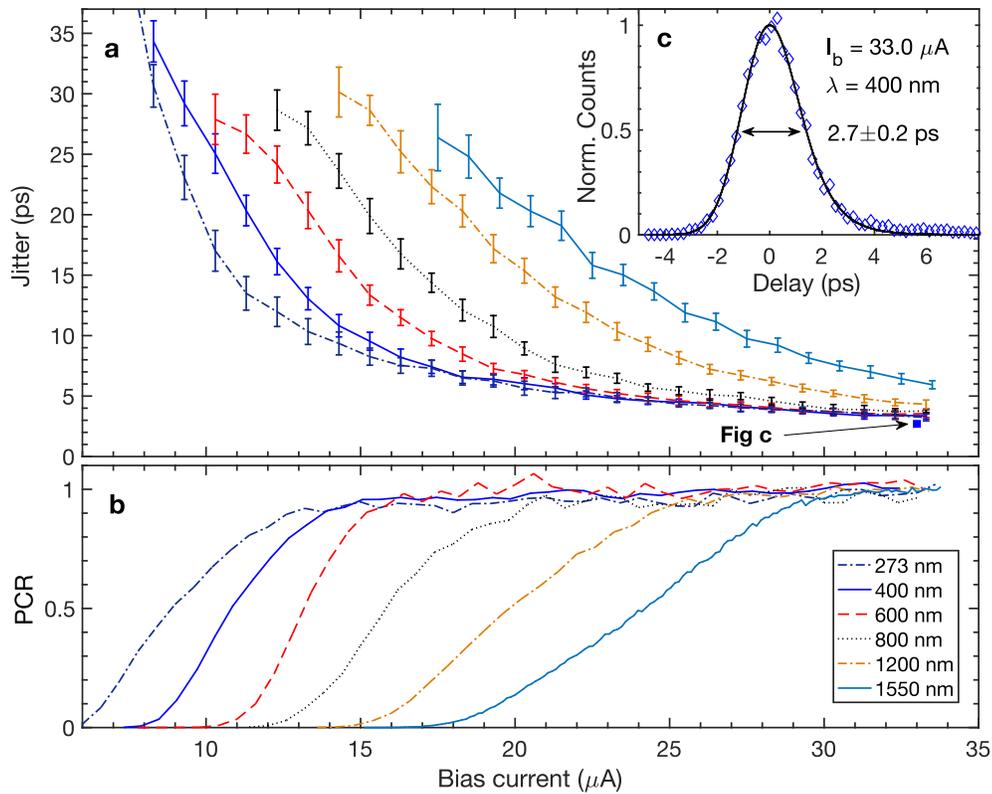

**Figure 4 | Wavelength dependence of the jitter.** Jitter (a) and normalized PCR (b) as a function of bias current for the 120 nm-wide nanowire measured with wavelengths of light between 273 nm and 1550 nm. We infer that the region where the jitter curves begin to overlap and saturate (above 17 μA for 273 nm and 400 nm and above 24 μA for 600 nm) is dominated by instrumental sources rather than intrinsic jitter. The error bars represent the 95% confidence bounds of the IRF fit (see Methods). (c) IRF for a bias current of 33 μA with 400 nm light, measured with an oscilloscope capable of double the sample rate (80 Giga samples per second) compared to the results presented in (a). The greater number of samples per waveform improves the accuracy of the delay measurement (see Methods), resulting in an improved jitter of 2.7±0.2 ps (compared to 3.2±0.2 ps obtained with the 40 Giga samples per second oscilloscope). We observed an anomalous change in the delay histogram tail at low bias currents for the ultraviolet light, which was not captured by the fitting function (below 0.25 of the peak value), see Supplementary Note 3 for details.

# SUPPLEMENTARY INFORMATION

**Supplementary Note 1 | Exponentially modified Gaussian function fitting parameters.** When fitting the exponentially modified Gaussian (EMG) function (See Methods in the main text), the shape of the distribution is described by two parameters; the width of the Gaussian distribution and the exponential distribution. Supplementary Figure 5(a) shows the full-width at half-maximum (FWHM) values of the fitted function for the four different nanowire widths studied, for the 775 nm and 1550 nm wavelengths (this figure is the same as Fig. 3(a) in the main text). Supplementary Figure 5(b) shows the FWHM value of the Gaussian fit parameter ($2.355\sigma$), and Supplementary Figure 5(c) shows the time constant of the exponential fit parameter ($1/\lambda$). Supplementary Figure 6 shows the same type of data for the 120 nm wide detector for wavelengths between 273 nm and 1550 nm (same data presented in Fig. 4(a) in the main text).

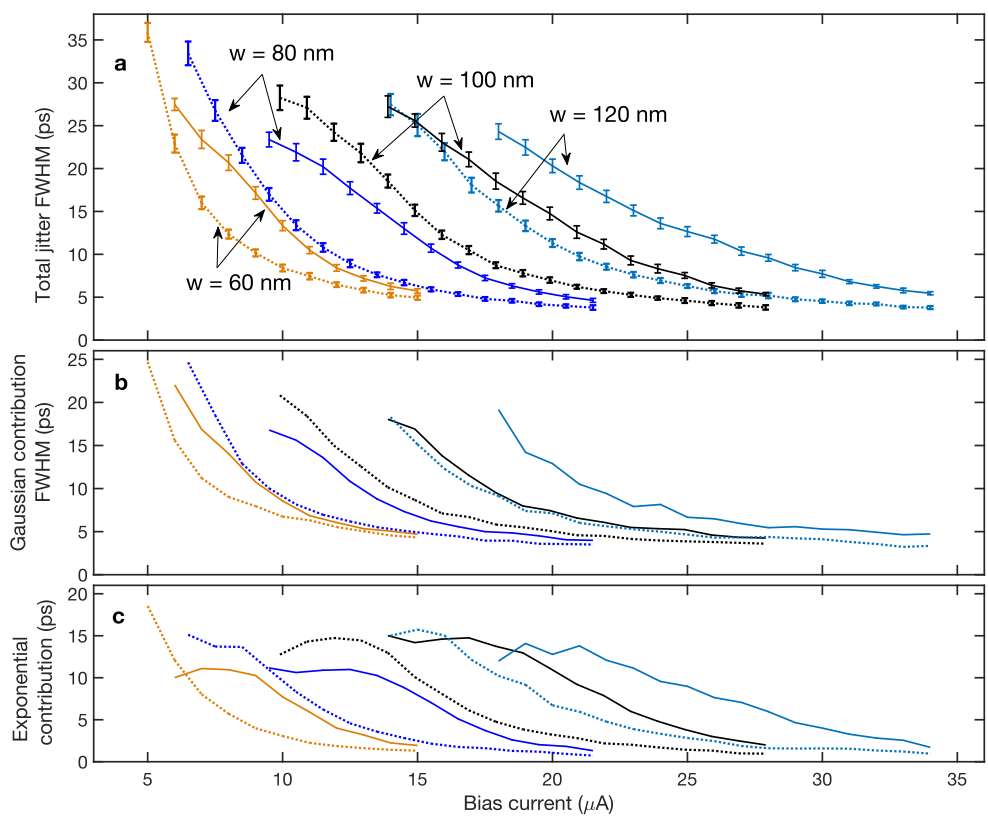

**Supplementary Figure 5.** Total temporal jitter (a) as well as the Gaussian (b) and exponential (c) fit parameters as a function of bias current, for different nanowire widths and illumination wavelengths. The nanowire widths are represented with different colors: 60 nm (orange), 80 nm (blue), 100 nm (black) and 120 nm (red). Illumination with 775 nm light is indicated by dotted lines and 1550 nm light by solid lines. The error bars in (a) represent the 95% confidence bounds of the IRF fit.

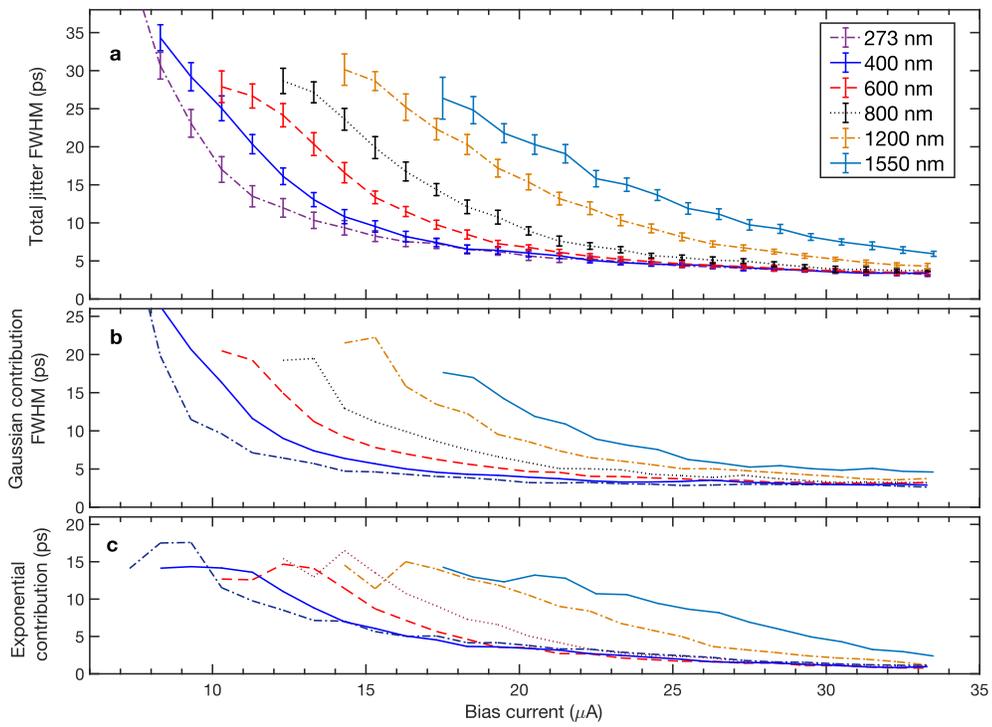

**Supplementary Figure 6.** Total temporal jitter (a) as well as the Gaussian (b) and exponential (c) fit parameters as a function of bias current, for the 120 nm nanowire illuminated with different wavelengths between 273 nm and 1550 nm. The error bars in (a) represent the 95% confidence bounds of the IRF fit.

**Supplementary Note 2 | Photon energy independence of SNSPD rising edge.** To ensure that the SNSPD signal shape remained independent of the illuminating photon energy we measured the signal slope at the trigger level for each bias current.

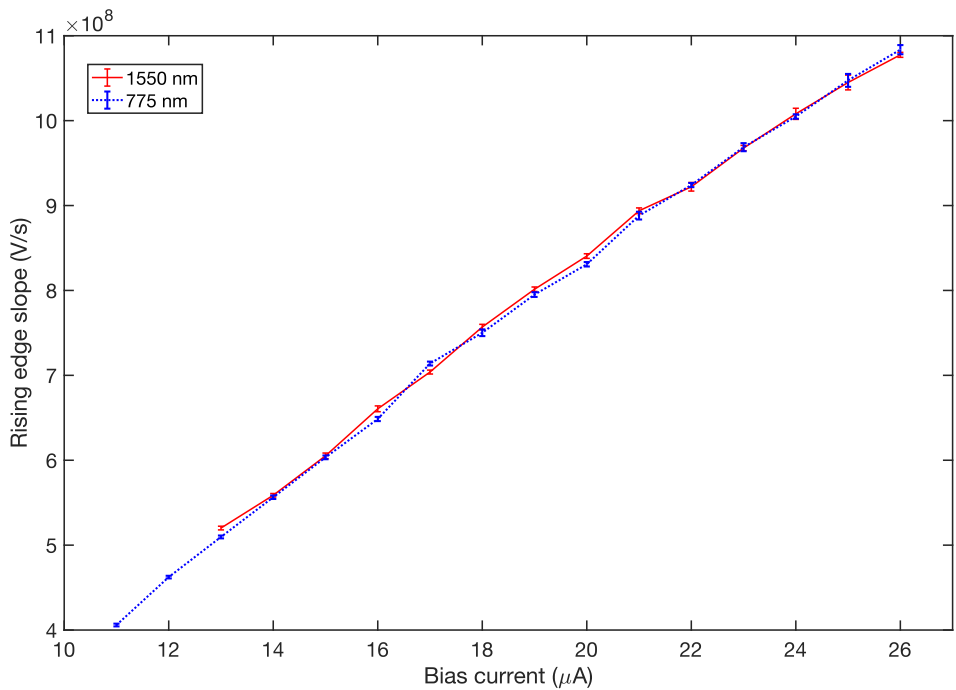

Supplementary Figure 7 shows an example of this measurement for the 100 nm-wide nanowire, where the rising edge slope is the same for both 775 nm and 1550 nm wavelengths. Note that both lines lie on top of each other.

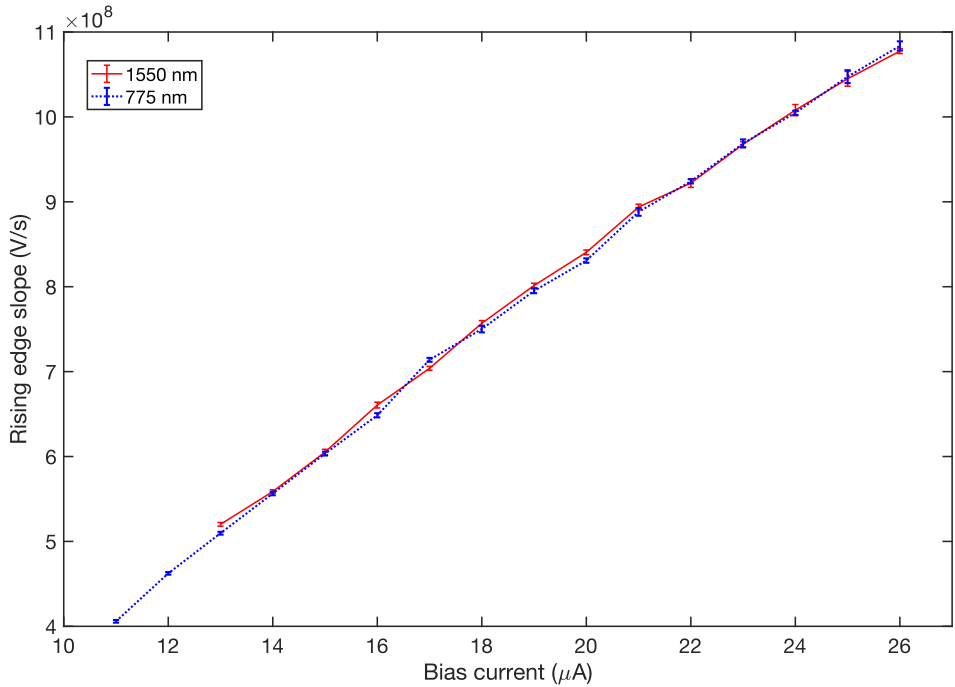

**Supplementary Figure 7.** SNSPD signal rising edge slope as a function of bias current for the 100 nm nanowire. Data for the 775 nm and 1550 nm illuminations shows that the pulse signal pulse shape is independent of photon energy. The slope change with bias current because the signal amplitude increases with the bias current flowing in the SNSPD.

**Supplementary Note 3 | Deviation from exponentially modified Gaussian (EMG) function for short wavelengths.** When characterizing the 120 nm-wide nanowire, we noticed that the tail of the distribution for the shortest wavelength seemed to have a different shape, which is not well represented by the EMG fitting function, especially at low bias currents. Supplementary Figure 8 shows a comparison of the instrument response functions for three wavelengths (273 nm, 400 nm and 800 nm) for different bias currents. At the lowest bias current of 15 µA, there is a systematic deviation from the fitting function

(below 0.25 of the peak) for the 273 nm wavelength data (Supplementary Figure 8(a)). This deviation decreases at longer wavelengths (Supplementary Figure 8(b,c)). The origin of this effect is unclear

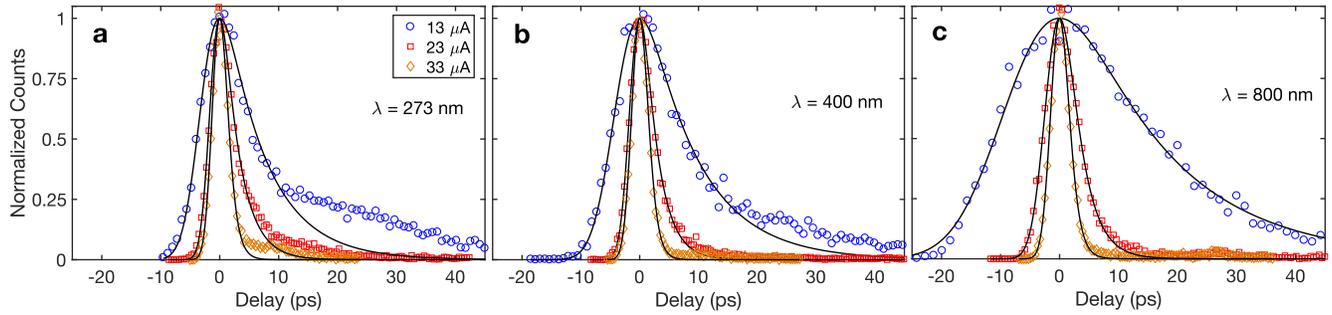

**Supplementary Figure 8.** Instrument response functions for the 120 nm nanowire biased at 13 µA (circles), 23 µA (squares) and 33 µA (diamonds). Data for the 273 nm (a), 400 nm (b) and 800 nm (c) is shown for comparison.

**Supplementary Note 4 | Temperature dependence of the jitter.** Supplementary Figure 9(a) shows the bias current dependence of the jitter at different temperatures between 0.9 and 4 K for the 80 nm-wide nanowire illuminated with 1550 nm light. At 4 K, a shift of about 10 sigma in the timing jitter was measured in the 11 – 13 µA range, which is also the range where the strongest wavelength dependence is observed in Fig. 3(a) in the main text. The switching current is suppressed above 14 µA at 4 K, making jitter measurements above this current impossible at this temperature. It was independently verified that

the shape of the electrical pulse and the level of the amplifier noise were independent of the operating temperature. Supplementary Figure 9(b) shows the photon count rates at the corresponding temperatures, which reflect the shift in the jitter curves. The presence of this reduction in the timing jitter at higher temperatures is an additional indication that intrinsic jitter mechanisms contribute to the timing uncertainty in this device.

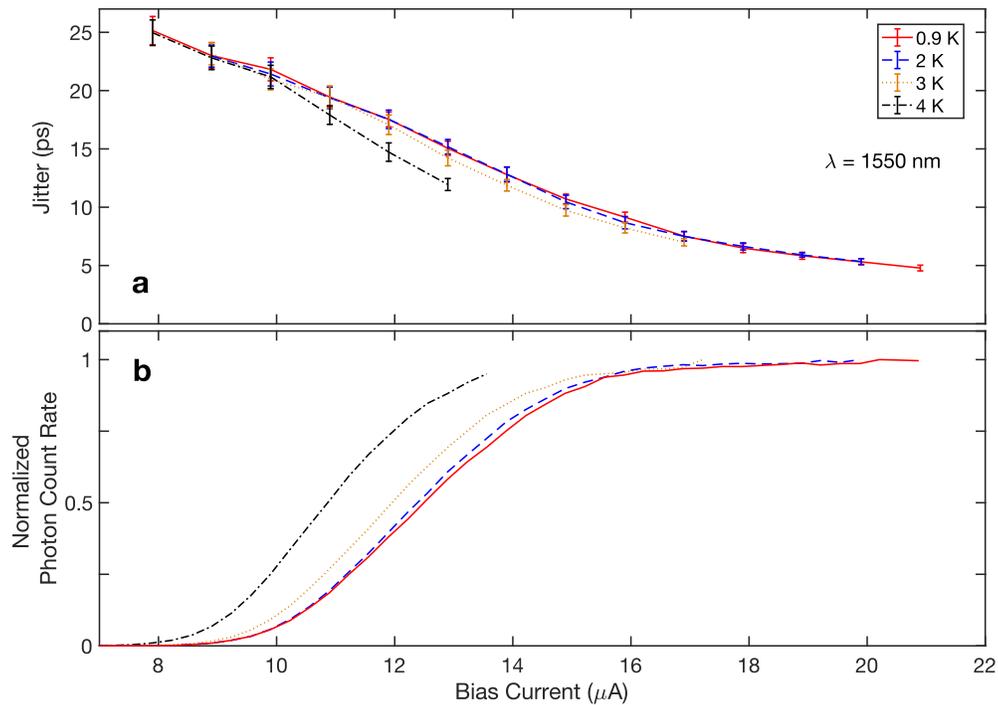

**Supplementary Figure 9.** Jitter (a) and normalized photon count rate (b) as a function of bias current for the 80 nm-wide nanowire measured with 1550 nm light at different operating temperatures between 0.9 K and 4 K. Because electrical-noise-limited jitter should not show a strong temperature dependence, the observed effect here suggests that another limiting mechanism (perhaps the intrinsic material properties) are dominating the observed jitter.